\documentclass{iau}
\usepackage{graphicx}

\usepackage[usenames]{color}

\title[Analyzing strange modes with TDC] 
      {Analysis of strange-mode instability \\
        with time-dependent convection \\
        in hot massive stars}

\author[Takafumi Sonoi \& Hiromoto Shibahashi]   
{Takafumi Sonoi \and Hiromoto Shibahashi}

\affiliation{Department of Astronomy, 
  University of Tokyo, Tokyo, 113-0033, Japan \\
  email: {\tt sonoi@astron.s.u-tokyo.ac.jp}, \ {\tt shibahashi@astron.s.u-tokyo.ac.jp} }

\pubyear{2013}
\volume{301}  
\pagerange{1--2}
\jname{Precision Asteroseismology}
\editors{J.A. Guzik, W.J. Chaplin, G. Handler \& A. Pigulski, eds.}
\begin{document}

\maketitle

\begin{abstract}
We carry out nonadiabatic analysis of 
strange-modes in hot massive stars with time-dependent convection (TDC) for the first time.
Although convective luminosity in envelopes of hot massive stars is not as dominative as in stars near the red edge of the classical Cepheid instability strip in the Hertzsprung-Russell (H-R) diagram, we have found that the strange-mode instability can be affected by the treatment of convection. However, existence of the instability 
around and over the Humphreys-Davidson (H-D) limit is independent of the treatment. This implies that the strange-mode instability could be responsible for the lack of observed stars over the H-D limit regardless of uncertainties on convection theories.  
\end{abstract}

``Strange'' modes with extremely high growth rates appearing in very luminous stars with $L/M\gtrsim 10^4L_{\odot}/M_{\odot}$ have significantly different characteristics from ordinary p- and g-modes (\cite[Wood 1976, Shibahashi \& Osaki 1981]{Wood1976, Shibahashi1981}).
Stability of 
strange-modes in hot massive stars has been 
analyzed with frozen-in convection (FC) by \cite{Glatzel1996}, \cite{Godart2011}, \cite{Saio2013}, and others.
In envelopes of hot massive stars, convective luminosity 
is not as dominative as in stars near
the red edge of the classical Cepheid instability strip. 
But the 
strange-modes are excited at the convection zones, and we cannot definitely conclude that effects of convection are negligible. 

\begin{figure}[t]
  \begin{minipage}[t]{0.49\textwidth}
    \centering
    \includegraphics[width=\hsize, height=0.63\hsize]{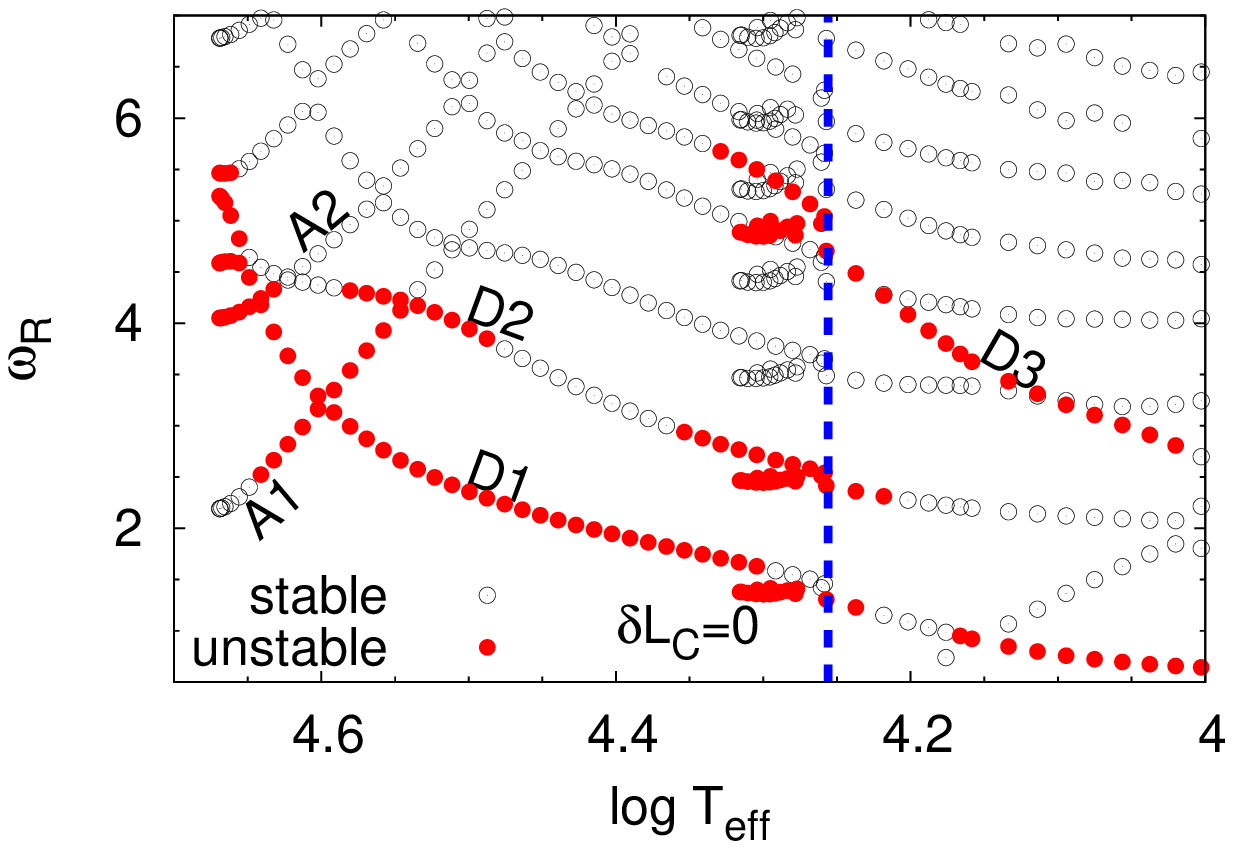}
  \end{minipage}
  \begin{minipage}[t]{0.49\textwidth}
    \centering
    \includegraphics[width=\hsize, height=0.63\hsize]{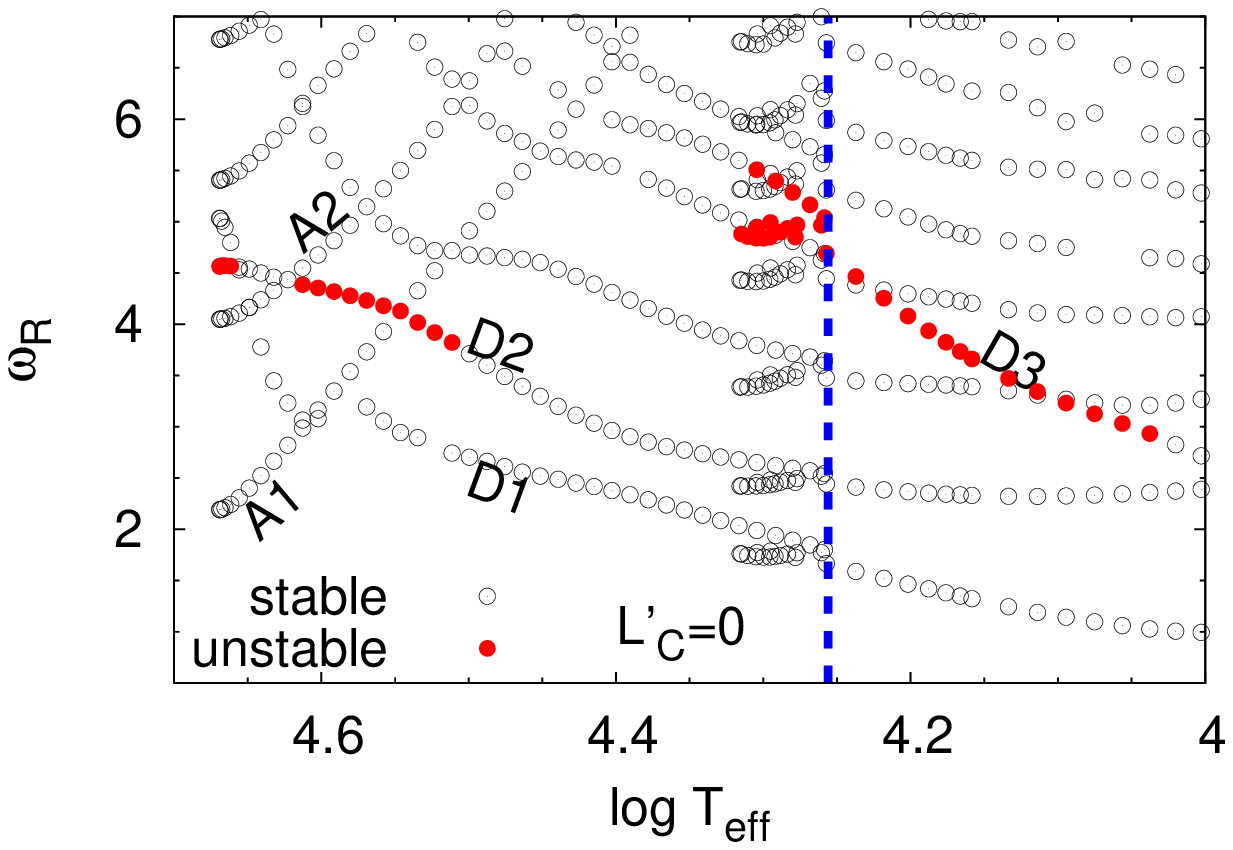}
  \end{minipage}
  \caption{Modal diagrams of radial modes for $60\,M_{\odot}$ with FC of zero Lagrangian ({\it left panel}) and Eulerian ({\it right panel}) perturbations of convective luminosity. The open and filled circles denote stable and unstable modes, respectively, and the vertical dashed line in each panel shows the cross point of the evolutionary track and the H-D limit.\label{fig:1}}
\end{figure}

We construct stellar models with $X=0.70$, $Z=0.02$ by using MESA (\cite[Paxton et al. 2011]{Paxton2011}), and analyze their radial modes with the nonadiabatic code developed by \cite{Sonoi2012}. We use the time-dependent convection (TDC) formulation 
derived by \cite{Unno1967} and developed later by \cite{Gabriel1974} and by \cite{Gabriel1996}. 
This TDC theory has already independently implemented by \cite{Grigahcene2005}.
Figure\,\ref{fig:1} shows the results for $60\,M_{\odot}$ with two types of FC; zero Lagrangian and Eulerian perturbations of convective luminosity, $\delta L_{\rm C}=0$ and $L'_{\rm C}=0$. The ascending sequences such as A1 and A2 correspond to ordinary modes, while the descending ones such as D1, D2 and D3 are 
strange-modes. In whole, instability is suppressed in the $L'_{\rm C}=0$ case compared with 
the $\delta L_{\rm C}=0$ case. The pulsations with $L'_{\rm C}=0$ are damped
due to the substantial gradient of convective luminosity near the boundaries of convective layers, while there is no convective damping with $\delta L_{\rm C}=0$. 

\begin{figure}[h]
  \begin{minipage}[t]{0.49\textwidth}
    \centering
    \includegraphics[width=\hsize, height=0.63\hsize]{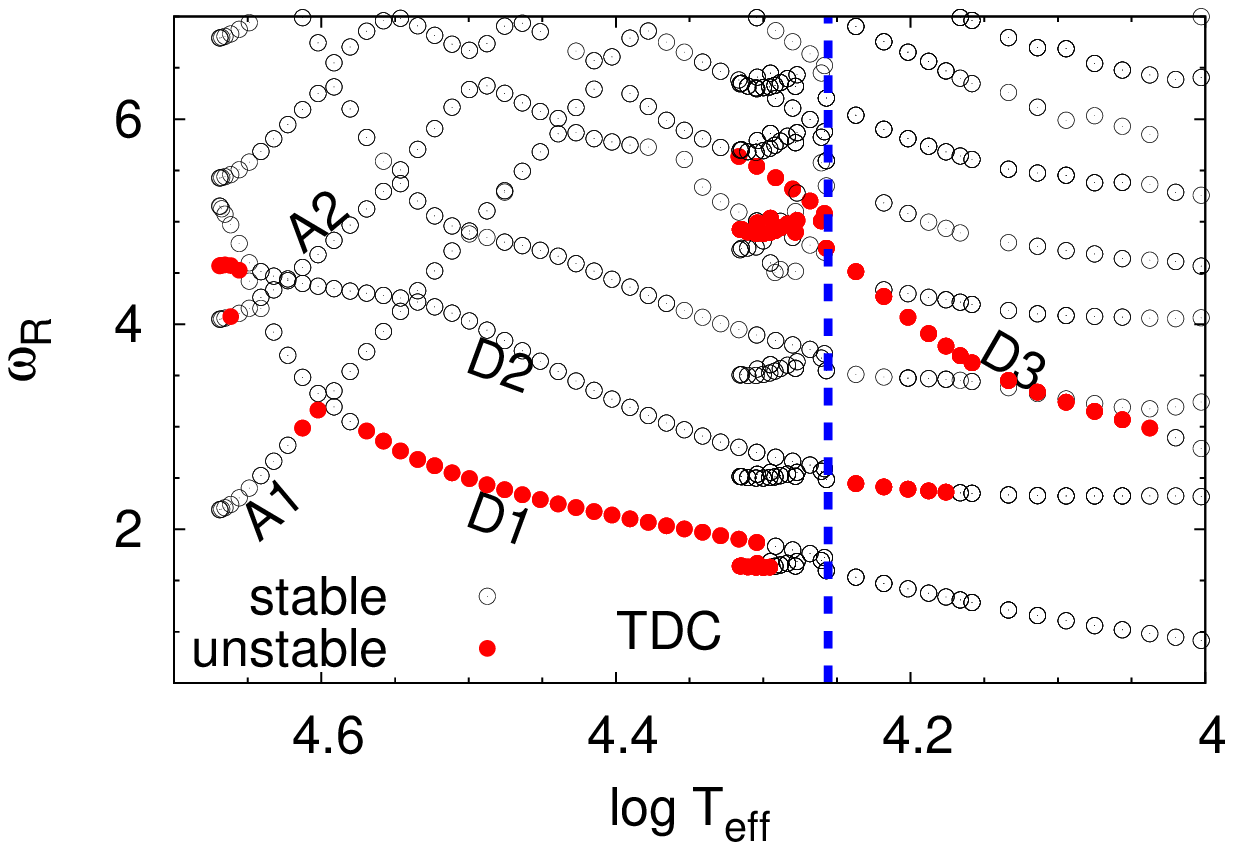}
  \end{minipage}
  \begin{minipage}[t]{0.49\textwidth}
    \centering
    \includegraphics[width=\hsize, height=0.63\hsize]{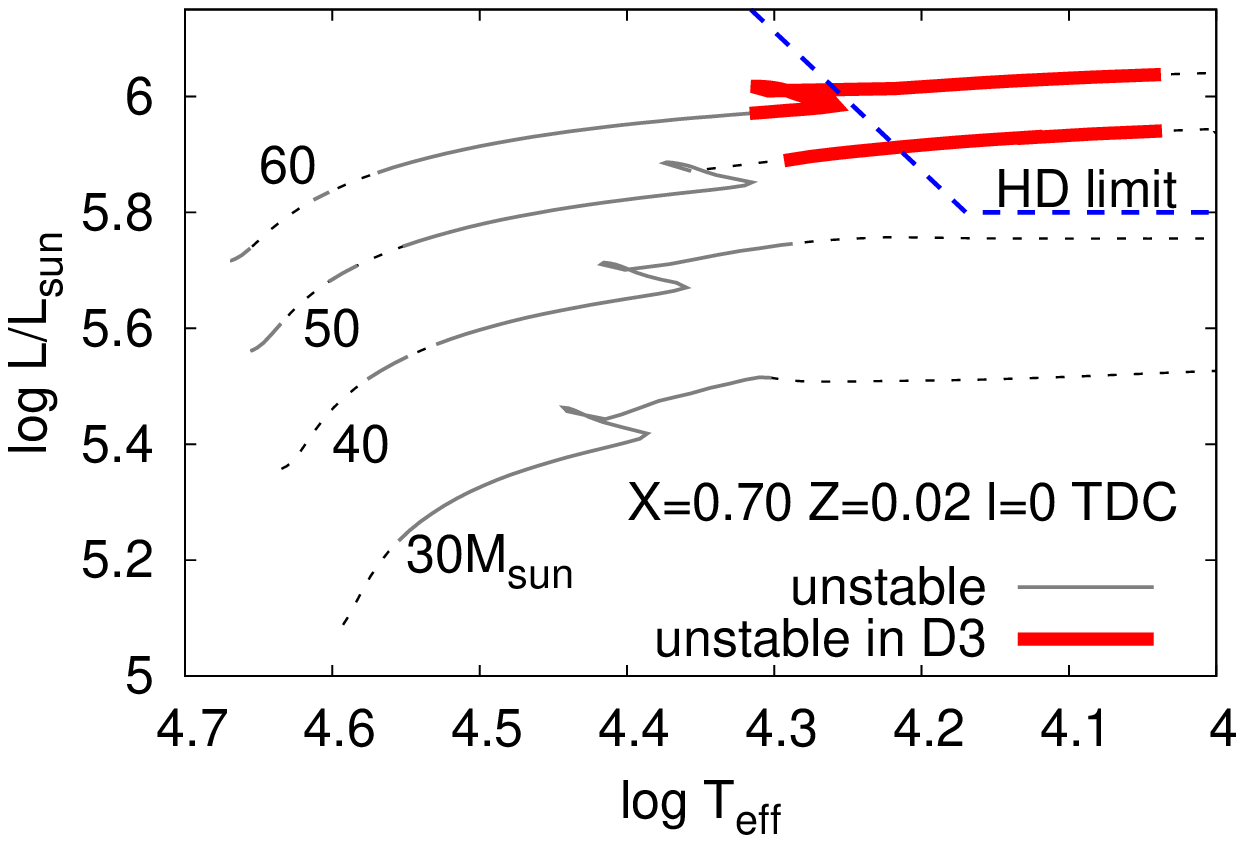}
  \end{minipage}
\caption{Results with TDC. Left: Same type of modal diagram as Fig. \ref{fig:1}. Right: Instability ranges on the evolutionary tracks derived with TDC on the H-R diagram. The thin dashed lines are evolutionary tracks, and the solid parts correspond to evolutionary stages having at least one unstable mode. Instability of D3 is particularly indicated with the thick solid lines.\label{fig:2}}
\end{figure}

Figure\,\ref{fig:2} shows the result with TDC. The degree of the instability is in between the two types of FC shown in Fig.\,\ref{fig:1}. 
While unstable modes of D1 and D2 are excited around the convective layer caused by the Fe opacity bump, where the ratio of the convective to the total luminosity $L_{\rm C}/L_{\rm r}$ is $\sim 10^{-1}$, 
the sequence of D3 is pulsationally unstable invariably in all the three treatments, as seen in Figs.\,\ref{fig:1} and \ref{fig:2}. 
These unstable modes of D3 are excited around the layer of He opacity bump, where $L_{\rm C}/L_{\rm r}$ is negligibly small, and hence we conclude that those modes are definitely unstable regardless of uncertainties of convection theories. 
We note that the instability of D3 appears around and above the H-D (\cite{HD1979}) limit. The resultant phenomena by the strange-mode instability have been proposed to be pulsationally driven mass loss by nonlinear analyses (e.g., \cite[Grott et al. 2005]{Grott2005}) and by observation (\cite[Aerts et al. 2010]{Aerts2010}). More detailed investigations are worth doing.

\end{document}